# High Performance Direct-Current Generator Based on Dynamic PN Junctions

*Yanghua Lu[1], Sirui Feng[1], Zhenzhen Hao[1], Runjiang Shen[1] and Shisheng Lin[1,2,]\**

1: College of Microelectronics, College of Information Science and Electronic Engineering, Zhejiang University, Hangzhou, 310027, China

2: State Key Laboratory of Modern Optical Instrumentation, Zhejiang University, Hangzhou, 310027, China

*Corresponding author. Tel: +86-0571-87951555

Email: shishenglin@zju.edu.cn.

## Abstract

After the electromagnetic generator, searching for novel electric generators without strong magnetic field is highly demanded. The generator without strong magnetic field calls for a physical picture distinct from the traditional generators. As the counterpart of the static PN junction has been widely used in the integrated circuits, we develop an electric generator named "dynamic PN generator" with a high current density and voltage output, which converts mechanical energy into electricity by sliding two semiconductors with different Fermi level. A dynamic N-GaAs/SiO$_2$/P-Si generator with the open-circuit voltage of 3.1 V and short-circuit density of 1.0 A/m$^2$ have been achieved. The physical mechanism of the dynamic PN generator is proposed based on the built-in electric field bounding back diffusing carriers in dynamic PN junctions, which breaks the equilibrium between drift and diffusion current in the PN junction. Moreover, the dynamic MoS$_2$/AlN/Si generator with the open-circuit voltage of 5.1 V and short-circuit density of 112 A/m$^2$ (11.2 mA/cm$^2$) have also been achieved, which can effectively output a direct-current and light up a blue light-emitting diode directly. This dynamic MoS$_2$/AlN/Si generator can continuously work for hours without obvious degradation, demonstrating its unique mechanism and potential applications in many fields where the mechanical energy is available.

## Main

The understanding of electron movements advances the information society and satisfying the requirements of electrical energy for human society. Michael Faraday is well known for the discovery of electromagnetic induction, who revealed the connection between magnetism and electricity in the year of 1831[1]. In electromagnetic generator, the electricity was generated when an electrical conductor passes through a magnetic field. Before that, electric generators were mainly referred to electrostatic generators under the operation principles of electrostatic induction and triboelectric effect[2]. Recently, the piezoelectric and triboelectric nanogenerators[3,4] attract many attentions, which could get electricity from the blue environmental energy such as mechanical and vibrating energy[5,6]. According to the proposed mechanism, the triboelectricity utilizes the displacement current in the Maxwell equation, which cannot flow freely through the insulating dielectric materials, may limiting its current output density[7]. On the other hand, solar cell works under the light illumination environments has although developed many novel structures[8-10]. It is imperative to search for a novel electric generator with high current density, which is the bottleneck of the current generators without high magnetic field under dark environments[11-16].

Since the invention of PN junction in 1940s[17], many applications have been explored such as the integrated circuits[18]. However, the dynamic PN junction has been rarely explored because most applications are focused on the static PN structure for the development of information society[19]. Recently, we have independently reported the dynamic Schottky diode as a direct current generator[20], and proposed the physical picture different from the triboelectricity[21,22]. However, as the Fermi level difference between metal and semiconductor is rather limited, which leads to a voltage output less than 1.0 V. The PN junction is totally different from Schottky diode as PN junction is minority carrier device while Schottky diode is majority carrier device, and the built-in voltage of the PN junction can be much larger than the Schottky diode as the Fermi level difference between N-type and P-type semiconductor can be several times of that of metal and semiconductor. Herein, we have demonstrated a high current density novel direct-current (d.c.) generator based on the dynamic PN junction, named as "dynamic PN generator", through sliding one semiconductor over another semiconductor with different Femi level, which can light up the light emitting diodes (LEDs) without any external rectifying circuits. The connected dynamic PN junction in series can output voltage larger than 9.0 V. The work mechanism is based on the built-in electric field separation of the otherwise diffused carriers as a result of the destruction and establishment of the depletion layer in dynamic PN junctions, which breaks the equilibrium between the drift and diffusion current, leading to an electricity output whose mechanism is different from other types of electric generators[1-4].

Typically, a dynamic GaAs/Si junction can convert the mechanical energy into electricity, whose open-circuit voltage ($V_{oc}$) reaches 0.7 V and short-circuit density ($J_{sc}$) reaches 1.8 A/m². Both the $V_{oc}$ and $J_{sc}$ are the peak value of voltage and current output. Furthermore, compared with the semiconductor-semiconductor dynamic PN generator, the semiconductor-insulator-semiconductor structure based dynamic PN generator outputs higher voltage attributed to the interface passivation and insulator bounding effect under the localized built-in electric field[23,24], whose $V_{oc}$ can reach 3.1 V and $J_{sc}$ can reach 1.0 A/m² for the case of GaAs/SiO$_2$/Si heterojunction. Moreover, a dynamic MoS$_2$/AlN/Si generator with $J_{sc}$ of 112 A/m² and $V_{oc}$ of 5.1 V has been achieved. The current density of 112 A/m² is orders of magnitude higher than triboelectric nanogenerators (~10³ times)[7,25,26] and piezoelectric nanogenerators (~10⁴ times)[7,27,28], which can be further enhanced through optimizing the PN junction interface and structure[26,29]. Without external rectifying circuit, this dynamic PN generator has the unique advantage of ultrahigh current density and direct-current output ability[20-22]. This dynamic PN generator can be connected in series or in parallel, which can effectively increase the output voltage or current. Two and three dynamic P-Si/N-GaAs generators in series can effectively output a voltage of 6.1 V and 9.0 V, which can light up a blue light-emitting diode (LED) directly. Especially, the dynamic P-Si/AlN/MoS$_2$ generator does not show obvious degradation after 60 min of running, confirming its mechanism is different from the triboelectricity and proving the potential applications of the dynamic PN generator[30,31].

As a representative dynamic PN generator, the schematic structure of the dynamic GaAs/Si generator is illustrated in **Figure 1**a. The N-type GaAs wafer was pressed on the P-type Si substrate compactly with a 6N force, which showed typical rectification behavior with a low leakage current density of 1.5 μA/cm$^2$ under the bias voltage of -3 V (Figure 1b). The work function of N-type GaAs and P-type Si used here is 4.10 eV and 5.12 eV (as shown in the Supplementary Note 1), respectively. Accordingly, the work function of N-GaAs is smaller than P-Si, thus a built-in electric field will be formed between the N-GaAs and P-Si substrate. Primarily, the N-type GaAs was dragged along the surface of the P-type Si and a maximal $V_{oc}$ up to 0.7 V was observed, which is independent on the working area. The dynamic P-Si/N-Si and P-Si/N-GaN junction can also output voltage signal as well as P-Si/N-GaAs junction, among which the $V_{oc}$ of P-Si/N-GaAs junction is the highest, which is 7 times of dynamic P-Si/N-Si generator (Figure 1c). The Fermi level of N-Si and N-GaN are about 4.28 eV and 4.35 eV (as shown in the Supplementary Note 1), so the generated voltage direction of P-Si/N-Si and P-Si/N-GaN generator is same as the P-Si/N-GaAs generator，confirming the importance of Fermi level (some other semiconductors are also explored in the Supplementary Figure 1). It is noteworthy that only noise electricity can be produced when sliding one piece of P-type Si over another P-type Si with same Fermi level (Supplementary Figure 2), indicating the important role of Fermi level difference and the built-in field. Figure 1d shows that the current output is positively proportional to the work area, indicating the current output can be improved by enlarging the work area. For a GaAs/Si dynamic PN generator with a working area of 1.0 mm$^2$ (2cm × 0.05mm), a maximal peak short-circuit current ($I_{sc}$) up to 1.8 μA was observed primitively, which was positive related to the movement speed within limits (Supplementary Figure 3). The $J_{sc}$ up to 1.8 A/m$^2$ can be achieved, which is orders of magnitudes higher than the reported polymer-based triboelectric nanogenerators (in the order of 10$^{-1}$ A/m$^2$)[7,25,26] and piezoelectric generators (in the order of 10$^{-2}$ A/m$^2$)[7,27,28]. Furthermore, the relationship between the $V_{oc}$/$I_{sc}$ and the force exerted on device is explored and it is found that the force of 6N is the optimal choice (Figure 1e), where the N-GaAs and P-Si can have the best rectification characteristic (as shown in J-V curves of the P-Si/N-GaAs junction under different forces in Supplementary Figure 4 and $N_{IF}$ in Supplementary Note 2). The 3D diagram for converting mechanical energy into electrical energy by the dynamic PN junction is shown in Figure 1f, which can work in the linearly reciprocating mode or circularly rotating mode. Figure 1g and Figure 1h show the $I_{sc}$ of the generator in linearly reciprocating mode and circularly rotating mode, respectively. It can be seen that the linear reciprocating mode outputs a pulsed current, which is caused by the inevitable acceleration and deceleration of the GaAs wafer motion. In contrast, when the GaAs wafer is continuously and evenly slide under the circularly rotating mode, it outputs a sustained direct current. The continuous d.c. output of our dynamic P-Si/N-GaAs generator indicates its unique physical mechanism different from the contact mode generator of PN generator or triboelectric generators.

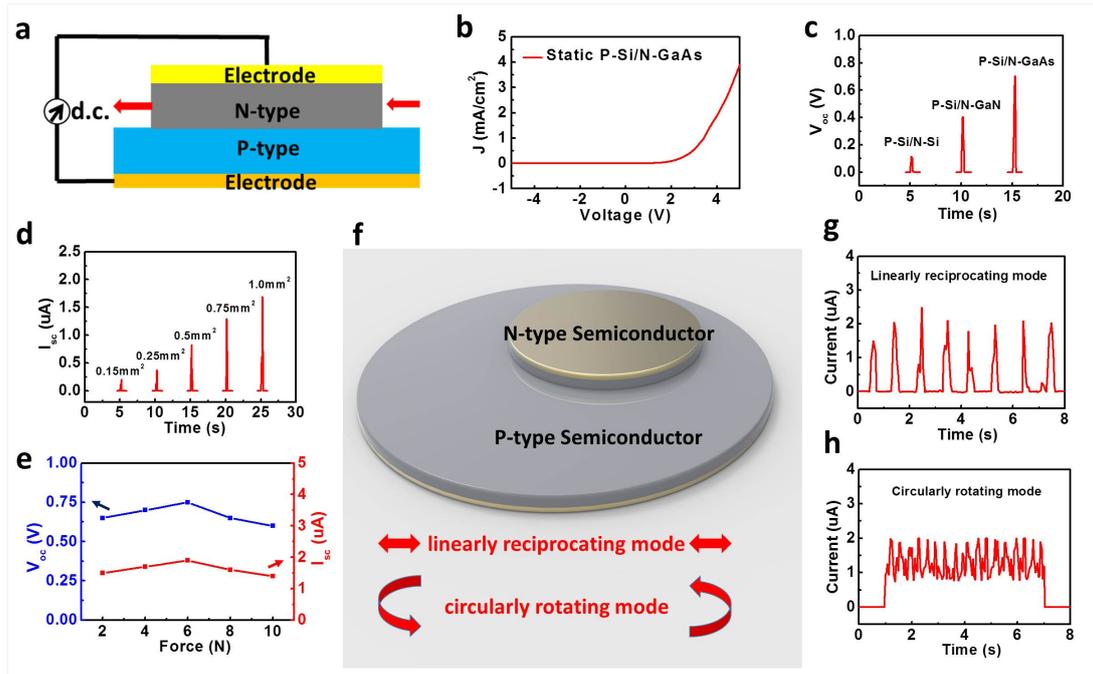

**Figure 1.** Experimental designs and results of the dynamic PN generator. a) The schematic illustration of the dynamic P-Si/N-GaAs generator. b) The J-V curve of the static P-Si/N-GaAs junction with a 6N force. The contact area is 1.0 mm$^2$. c) The $V_{oc}$ of dynamic P-Si/N-Si, P-Si/N-GaN, P-Si/N-GaAs generator with a speed of 6cm/s and a 6N force. d) The $I_{sc}$ of dynamic P-Si/N-GaAs generator with a 6N force, a speed of 6 cm/s and the work area of 0.15, 0.25, 0.5, 0.75, 1.0 mm$^2$. e) The $V_{oc}$ and $I_{sc}$ of dynamic P-Si/N-GaAs generator with different force exerted on the junction in a speed of 6 cm/s. f) Experimental design and 3D diagram for converting mechanical energy into electrical energy by a dynamic PN junction. The inset shows the linearly reciprocating and circularly rotating modes of the generator, respectively. g) and h) show the $I_{sc}$ of dynamic P-Si/N-GaAs generator under the linearly reciprocating mode and circularly rotating mode with a 6N force and a speed of 6 cm/s, respectively.

As schematically shown in **Figure 2**, when we move the PN junction in time pieces, the destruction of the PN junction at the rear part and establishment at the front part will happen subsequently. We name this junction 'the dynamic PN junction'. The destruction of the PN junction at the rear part inhibits the diffusion paths for the diffusing electrons and holes, which is necessary for balancing the drift current. Those otherwise diffused carriers are bound back and accelerated by the built-in field. As the P-type side contacts with N-type side at the front part, the drift current will balance with the diffusion current very soon. As time passes in a series of pieces, more electrons and holes are bound back the N-type side and P-type side, respectively, leading a net current output and building a voltage output. Figure 2a shows the schematic illustration of the P-Si/N-GaAs interface. Actually, for the P-Si/N-GaAs junction, the drift-diffusion current equation can be described as below[32]:

$$J_n = J_n^{drift} + J_n^{diff} = qnu_n E + qD_n \nabla n \qquad (1)$$

$$J_p = J_p^{drift} + J_p^{diff} = qpu_p E - qD_p \nabla p \qquad (2)$$

where $J_n$, $J_p$, $\mu_n$, $\mu_p$, $D_n$, $D_p$ are the electron/hole current density, the electron/hole mobility and the electron/hole diffusion coefficient of GaAs and Si, respectively. $E$ is the built-in field, $q$ is the elementary charge, $n$ and $p$ are the position dependent electron/hole density in GaAs and Si, respectively. When the GaAs wafer moves along the Si substrate, there is a dynamic process of disappearance of the depletion layer in the rear end and the re-establishment of the depletion layer in the front end. In macro scale, the effective working area of the dynamic PN junction is assumed as the contact area of the GaAs wafer and Si substrate during the movement. According to equation (1) and (2), the current density of GaAs/Si junction is consisted of $J^{diff}$ and $J^{drift}$. However, the dynamic process of the generation and disappearance of the depletion layer in dynamic GaAs/Si junction will break up the static carrier distribution equilibrium and bound back the electrons or holes which are diffusing across the depletion layer. The bounding back electrons and holes under the built-in field lead to increase of $J^{drift}$ and decrease of $J^{diff}$, breaking up the equilibrium of drift-diffusion current in the static PN junction, which is the origin of the output current generated in dynamic PN junction. The rectification characteristic of the dynamic N-GaAs/P-Si junction from -5V to 5V was measured and shown in Figure 2b, where the oscillation of the J-V curve as voltage increases. The oscillation of the current value indicates there are destruction and re-establishment of the depletion layer in the dynamic PN junction, which is in agreement with the previous proposed physical picture. The detailed transportation of the bound back electron and hole are shown in Figure 2c, where also includes the band gap alignment between P-Si and N-GaAs. The built-in field point to P-Si as a result of electron diffusion from N-GaAs to P-Si. When the N-GaAs wafer is dragged on P-Si, the diffusing electrons and holes in the rear part has no way to diffuse into P-Si and N-GaAs, respectively. Thus, under the effect of built-in field, the otherwise diffused electron and holes in the rear part are bound back to the N-GaAs and P-Si, respectively, breaking up the equilibrium of drift-diffusion current and forming electrical output. So the built-in electric field and the barrier height of the PN junction must play a crucial role. The temperature difference of semiconductors before and after the friction is very limited

(Supplementary Figure 5), which indicates that the heat effect is ignorable for outputting electrical signal.

To reveal this distinctive current generation mechanism and prove the barrier height of the PN junction plays a crucial role in our generator, P-type silicon wafers with different thickness (d) of silicon oxide are used here for further experiments. We find this dynamic semiconductor-insulator-semiconductor structure generator has higher voltage output because of the increased barrier height[23]. As shown in band diagram of the dynamic N-GaAs/SiO$_2$/P-Si junction (Figure 2d), the barrier height for bounding back electrons and holes are largely increased as the conduction/valence bands of SiO$_2$ are much higher/lower than that of GaAs and Si. When we move the N-GaAs wafer along the SiO$_2$/P-Si substrate, the diffusing electrons and holes can also be switched back and accelerated by the localized built-in electric field, moving toward the ohmic contact of N-GaAs and P-Si, respectively. The inserted SiO$_2$ layer can largely suppress the transfer of electrons and holes, stimulating the bounding back of diffusing electrons and holes. With the increase of barrier height, more bounding back electrons and holes accumulate in two semiconductors, increasing the potential difference as well as the voltage output. Figure 2e indicates that the rectification characteristic can be largely optimized after inserting the insulating layer, indicating the increase of the barrier height of the GaAs/Si junction. As shown in Figure 2f, the $V_{oc}$ and $J_{sc}$ of the dynamic P-Si/N-GaAs junction with the SiO$_2$ thickness of 0/9/13/20/50/100/200 nm are 0.7/0.9/1.2/1.6/2.2/3.1/4.0 V and 1.8/1.6/1.5/1.4/1.2/1.0/0 A/m$^2$, respectively. With the increase of the thickness of SiO$_2$,

the $J_{sc}$ decreases faintly to zero for 200 nm thick of $SiO_2$ but the $V_{oc}$ increases continuously. It is proved that the 200 nm $SiO_2$ is too thick for carriers crossing over it (Supplementary Figure 6), which lead to a nearly zero current output. The voltage output can be larger than the Fermi level differences for the cases of thick $SiO_2$, which should be ascribed to pumped electrons and holes in the N-type and P-type semiconductor, as the thick $SiO_2$ provides a high and safe barrier for bounding back the otherwise diffused electrons and holes compared with the case of thin $SiO_2$.

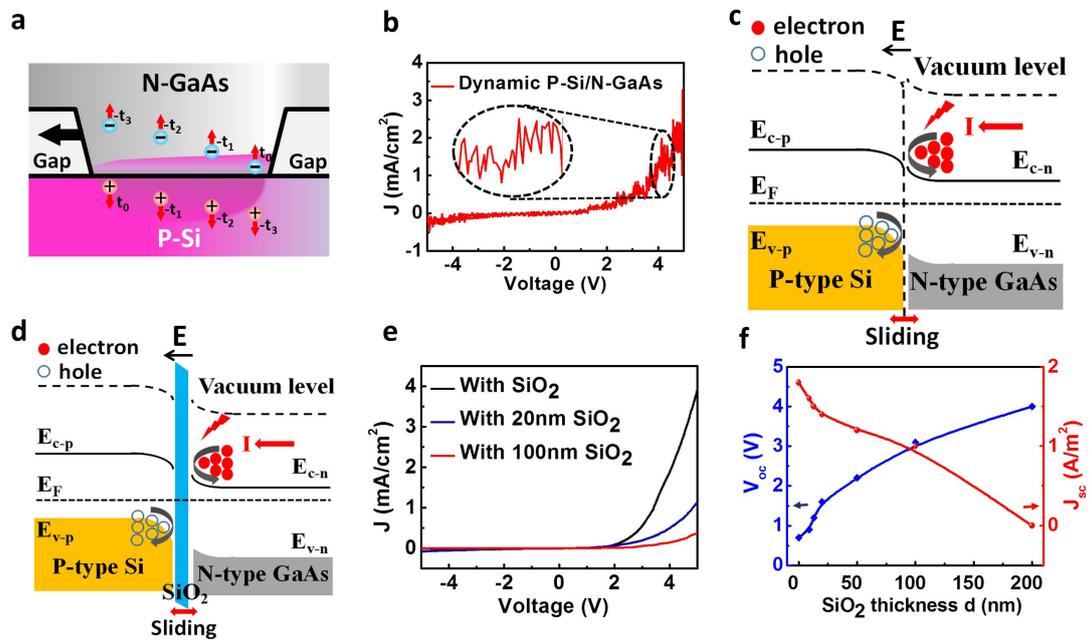

**Figure 2. a) The schematic diagram of the dynamic P-Si/N-GaAs generator. b) The J-V curve of the dynamic P-Si/N-GaAs junction with a 6N force. The contact area is 1.0 mm². c) The band diagram and carrier dynamic process of the dynamic P-Si/N-GaAs generator and d) the dynamic P-Si/SiO₂/N-GaAs generator. e) J-V curves of the dynamic P-Si/N-GaAs generator without SiO₂ and with 20/100 nm SiO₂. f) The $V_{oc}$ and $J_{sc}$ of the dynamic P-Si/N-GaAs generator with a 6N force, a speed of 6 cm/s and SiO₂ thickness of 0/9/13/20/50/100/200 nm.**

**The work area of the P-Si/N-GaAs junction is 1.0 mm$^2$. Both the $V_{oc}$ and $J_{sc}$ are the peak value of voltage and current output.**

To explore the potential practical applications, we charged a capacitor C (0.1 µF) by manually sliding a GaAs wafer on the SiO$_2$/Si substrate. Figure 3a and Figure 3b show the voltage of charging capacitor in the linearly reciprocating mode and circularly rotating mode, respectively. It can be seen that the linear reciprocating mode outputs a pulsed voltage and the circularly rotating mode outputs a continual voltage as high as 3.1 V under the consistent charging. In addition, the $V_{oc}$ outputs of the generator in parallel, 2-series and 3-series are 3.1/6.1/9.0 V, indicating the linear superposition effect of our devices, which can expand its applications (Figure 3c). The captured optical pictures of this capacitor charging experiment is shown in Figure 3d, which is consist of a capacitor (0.1 µF), a resistance (47 kΩ), a Keithley 2010 system and a reset switch. The detailed circuit diagram of the capacitor charging system and partial enlarged image of this dynamic P-Si/SiO$_2$/N-GaAs generator are shown in Figure 3e and 3f. It can be seen that no additional rectification was used. Therefore, this simple dynamic P-Si/SiO$_2$/N-GaAs generator has huge possibility for practical applications, which can transfer the mechanical energy into direct current output directly.

On this basis, we use the electricity of our devices to light up a blue LED (the picture of experimental model is shown in Supplementary Figure 7). This LED can only be light up under the drive voltage more than 2.7 V. Each of generators can stably provide nearly 3.1 V open-circuit voltage and 1.0 μA short-circuit current output. This dynamic PN generator can be connected in series or in parallel, which can effectively increase the output voltage or current. So two or three generators can ensure to light up the blue LED here (the circuit diagram of the LED lighting experiment is shown in figure 3g). Two generators in series can effectively output a voltage of 6.1 V, which can be used to provide enough voltage output for lighting a blue LED. Figure 3h shows the picture of the LED powered by our dynamic PN generator. It can be observed the blue light LED is lighted simultaneously with the lateral movement of PN junctions, indicating the output of a continuous direct current (the right inset).

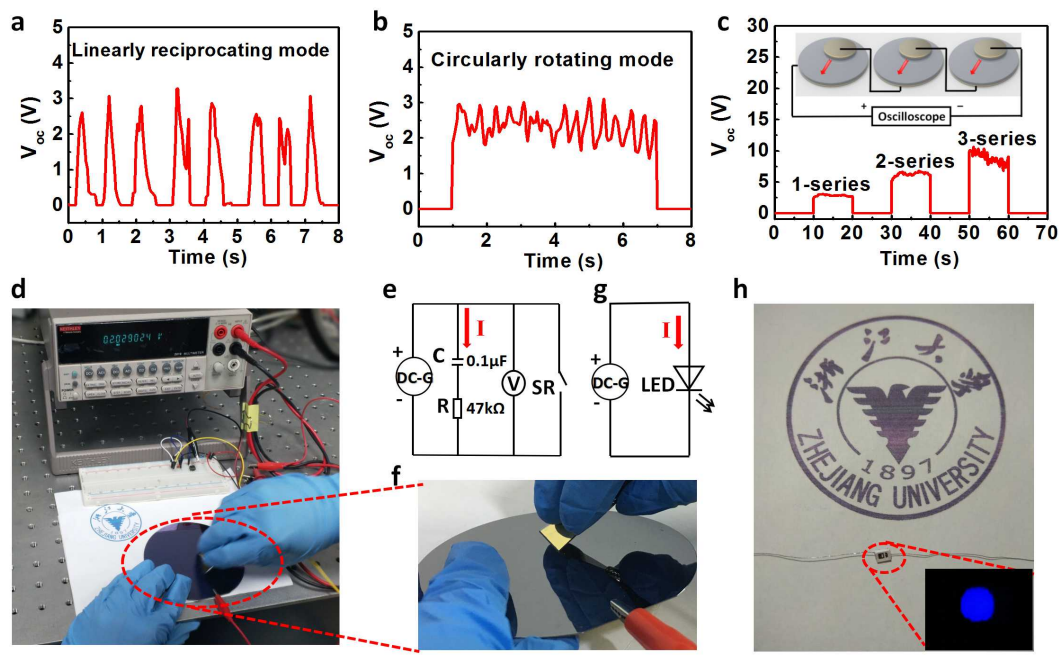

Figure 3. The $V_{oc}$ of dynamic P-Si/100nm-SiO$_2$/N-GaAs generator under a) the linearly reciprocating mode and b) the circularly rotating mode. c) The $V_{oc}$ of the dynamic P-Si/100nm-SiO$_2$/N-GaAs generator in 1-series, 2-series and 3-series

**with a 6N force, a speed of 6 cm/s. Inset: The schematic image of series connection of the dynamic PN junctions. d) The captured optical picture of the capacitor charging experiment, which is measured with Keithley 2010 system. e) The circuit diagram of charging a capacitor C (0.1μF) with the dynamic P-Si/SiO$_2$/N-GaAs generator. No additional rectification circuit has been used. f) The partial enlarged image of the dynamic P-Si/SiO$_2$/N-GaAs generator. g) The circuit diagram of the LED lighting experiment. h) Pictures taken from video to show the luminance of a LED powered by our dynamic PN generator. Right inset: The lighting blue LED powered by the dynamic PN junctions.**

As the friction strength between P-Si and N-GaAs is high, which may destroy the surface of the semiconductor, we have used the layered semiconductors such as MoS$_2$[21,33]. The diffusing electrons and holes in dynamic P-Si/10nm-AlN/MoS$_2$ junction can also be directionally bound back by the built-in electronic field at the interface, generating the electricity. As shown in the band diagram of dynamic P-Si/AlN/MoS$_2$ junction (Figure 4a), under the localized built-in electric field from MoS$_2$ to P-Si substrate, the diffusing electrons will bound back to the MoS$_2$ and diffusing holes will bound back to the P-type Si, breaking up the equilibrium of drift-diffusion current and forming electrical output. The rectification characteristic of the dynamic P-Si/AlN/MoS$_2$ junction from -5V to 5V is also measured. As shown in Figure 4b, the interfacial PN junction varies when the P-Si/AlN/MoS$_2$ contact is under the friction process, which helps testifying the theory proposed above. The fluctuation of the J-V curve of the P-Si/AlN/MoS$_2$ junction under friction condition is caused by the destruction and re-establishment of the depletion layer in the dynamic PN junction, which indicates the generation of the voltage output. Normally, the J$_{sc}$ as high as 112 A/m$^2$ can be achieved, with the work area of 0.05 mm$^2$, which is about three orders of magnitude higher than the reported triboelectric nanogenerators[25-28]. To demonstrate the power output of our generator, the voltage and current output as a function of electrical load R have been investigated and shown in Figure 4c. An increasing voltage and decreasing current density can be measured with the increase of load resistance. Accordingly, the power density output as a function of electrical load R have also been investigated and shown in Figure 4d. The peak power density about 130 W/m$^2$ can be found around R equals to 360 kΩ, which is close to the internal resistance of the PN junction. And the energy-conversion efficiency of the moving Si/MoS$_2$ junction generator can be calculated to be about 27.1% (as shown in the Supplementary Note 3). As MoS$_2$ is a lubricating material, dynamic P-Si/AlN/MoS$_2$ PN junction generator shows excellent stability. Notably, as shown in the direct and continuous voltage and current output of the dynamic P-Si/AlN/MoS$_2$ (Figure 4e and 4f), this dynamic P-Si/AlN/MoS$_2$ generator does not show any obvious degradation after 60 min of running, demonstrating the potential applications of the dynamic PN generator. The fluctuation of the continuous voltage and current output is caused by the vibration of the pressure and moving speed.

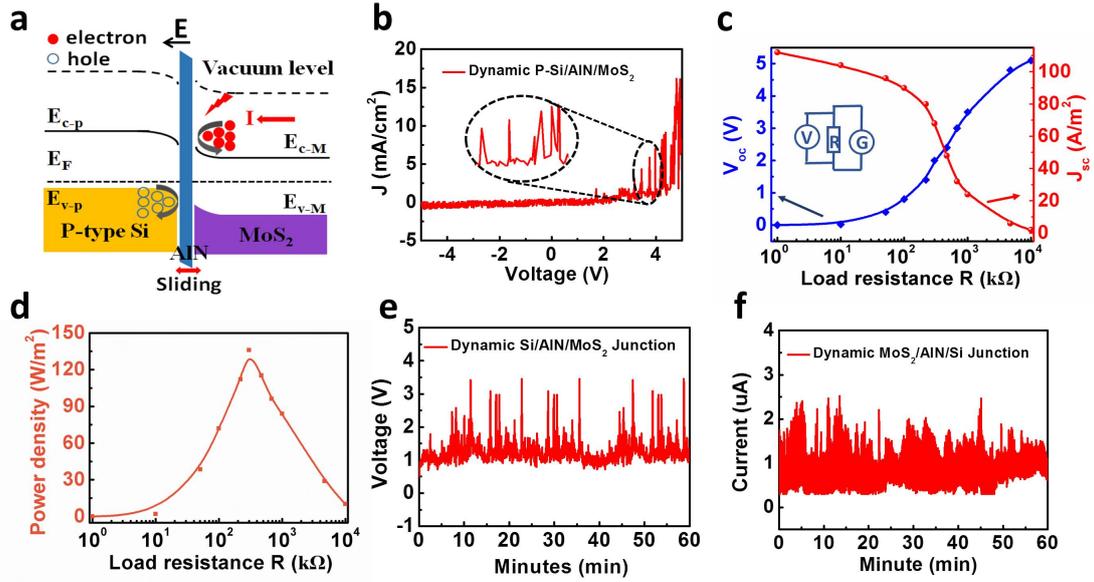

**Figure 4.** a) The band diagram and carrier dynamic process of the dynamic P-Si/AlN/MoS$_2$ generator. b) The J-V curve of the dynamic P-Si/AlN/MoS$_2$ junction with a 6N force. The contact area is 0.05 mm$^2$. c) V$_{oc}$ and J$_{sc}$ output as a function of electrical load R. Both the V$_{oc}$ and J$_{sc}$ are the peak value of voltage and current output. Insert: equivalent circuit. d) The power density output as a function of electrical load R. e) The direct and continuous voltage and f) current output of the dynamic P-Si/AlN/MoS$_2$ generator for 60 min continuously, with an electrical load R of 680 kΩ.

We have demonstrated a novel direct-current generator based on the dynamic PN junction, where two semiconductors with different Fermi level slide with each other. The mechanism is proposed based on the built-in electric field bounding back diffusing carriers emitted by the destruction and re-establishment of the depletion layer in dynamic PN junctions, which breaks the equilibrium between the drift current and diffusion current. The voltage, current and output power of the dynamic PN generator can be further enhanced through optimizing the PN junction interface and designing the semiconductor-insulator-semiconductor structure. Notably, the dynamic PN generators with high voltage of 5.1V and high current density of 112 A/m$^2$ have been realized based on MoS$_2$/AlN/Si structure. Especially, this dynamic P-Si/AlN/MoS$_2$ generator does not show obvious degradation after 60 min of running, demonstrating the unique mechanism and stability of the dynamic PN generator. This novel dynamic PN generator with sufficient current density has many promising applications in many fields where the mechanical energy is available.

**METHODS**

**Devices fabrication.** Firstly, the double side polished N-type GaAs wafer was dipping into 10 wt% HCl for 10 min to remove the native oxide layer in the surface and washed by deionized water. Then Au (100 nm) electrode was fabricated with magnetron sputtering on one side of GaAs wafer. Similarly, the single side polished P-type and N-type Si substrate were dipping into 10 wt% HF for 10 min to remove the native oxide layer in the interface and then washed by deionized water. Ti/Au (20 nm/50 nm) electrode was fabricated with magnetron sputtering on the unpolished side of Si substrate and MoS$_2$ flake. The N-type GaAs or MoS$_2$ was pressed closely on the P-type Si substrate by the hand, making sure a solid electrical contact between N-type GaAs/MoS$_2$ and P-type Si substrate can be achieved. The AlN layer on P-type Si is fabricated with Physical Vapor Deposition method. And the SiO$_2$ layer on P-type Si is fabricated with thermal oxidation method in 1100℃.

**Physical characterization methods.** The microscopic image of the generator was characterized with ZEISS optical microscopy. The current-voltage (I-V) curve of the PN junction was measured with Keithley 2400 system. The real-time voltage and current output were recorded in real time by a Keithley 2010 system, which was controlled by a LabView-based data acquisition system with a sampling rate of 25 s$^{-1}$. The Keithley 6485 picoammeter was also used to verify accuracy of current output, which was controlled by a LabView-based data acquisition system with a sampling rate of 100 s$^{-1}$. The force was measured with the pressure meter.

## Data availability

The data that support the findings of this study are available from the corresponding author upon reasonable request.

**Acknowledgements:** S. S. Lin thanks the support from the National Natural Science Foundation of China (No. 51202216, 51502264 and 61774135) and Special Foundation of Young Professor of Zhejiang University (Grant No. 2013QNA5007).

**Author Contributions:** S. S. Lin designed the experiments, analyzed the data and conceived all the works. Y. H. Lu carried out the experiments, discussed the results and wrote the paper. S. R. Feng, Z. Z. Hao and R. J. Shen discussed the results and assisted with experiments. All authors contributed to the writing of the paper.

**Author Information:** Reprints and permission information is available online. The authors declare no competing financial interest. Readers are welcome to comment on the online version of the paper.

**Supplementary Information** is Available in the online website.